\documentclass{aa}
\usepackage{graphicx}
\begin{document}
   \title{Images for an Isothermal Ellipsoidal Gravitational 
Lens from a Single Real Algebraic Equation} 

   \author{Hideki Asada
           \inst{1,2}, 
           Takashi Hamana
           \inst{3}, and 
           Masumi Kasai
           \inst{1}
          }

   \offprints{H. Asada\\ email: asada@phys.hirosaki-u.ac.jp}

   \institute{Faculty of Science and Technology, 
              Hirosaki University, Hirosaki 036-8561, Japan 
        \and 
              Max-Planck-Institut f\"ur Astrophysik, 
              Karl-Schwarzschild-Str. 1, 
              D-85741 Garching, Germany 
        \and 
              National Astronomical Observatory, 
              Mitaka 181-8588, Japan}

   \date{Received ; accepted }

   \abstract{
We present explicit expressions for the lens equation 
for a cored isothermal ellipsoidal gravitational lens as 
a a single real sixth-order algebraic equation in two approaches; 
2-dimensional Cartesian coordinates and 3-dimensional polar ones.
We find a condition for physical solutions which correspond to 
at most five images. For a singular isothermal ellipsoid, 
the sixth-order equation is reduced to fourth-order one for which  
analytic solutions are well-known. Furthermore, we derive 
analytic criteria for determining the number of images 
for the singular lens, which give us simple expressions for 
the caustics and critical curves. 
The present formulation offers a useful way for studying 
galaxy lenses frequently modeled as isothermal ellipsoids.

   }

\authorrunning{Asada, Hamana and Kasai}

\titlerunning{Isothermal Ellipsoidal Gravitational Lens}

   \maketitle
%

\section{Introduction}
Gravitational lensing due to a galaxy is important for 
probing mass distributions and determining cosmological
parameters. Galaxy lenses are often modeled as cored 
isothermal ellipsoids. Although the ellipsoidal model is 
quite simple, it enables us to understand a number of 
physical properties of the galactic lens. Furthermore, it 
fits well with mass profiles implied by observations 
(For instance, Binney and Tremaine 1987). 

Until now, the lens equation for the cored isothermal ellipsoid 
has been solved numerically as a {\it nonlinearly coupled} system. 
For a binary gravitational lens, it has recently been shown 
that the lens equation is reduced to a single real fifth-order 
algebraic equation (Asada 2002). 
Also for a singular isothermal ellipsoidal lens, furthermore, 
the apparently coupled lens equations can be reduced to 
a single equation (Schneider et al. 1992), 
though explicit expressions were not given there. 
Along this course, we reexamine the coupled lens equations 
for the cored isothermal ellipsoid. 
The main purpose of the present paper is to show that 
they are reduced to a single equation with a condition for 
physical solutions and to give analytic criteria for determining 
the number of images for isothermal ellipsoidal lenses.

\section{Lens Equation for a Singular Isothermal Ellipsoid} 
First, let us consider the singular isothermal ellipsoidal lens 
with ellipticity $0\leq\epsilon<1/5$. A condition that 
the surface mass density projected onto the lens plane 
must be non-negative everywhere puts a constraint 
on the ellipticity as $\epsilon<1$. 
A tighter constraint $\epsilon<1/5$ comes from that the density 
contours must be convex, which is reasonable 
for an isolated relaxed system. 
The lens equation is expressed as 
\begin{eqnarray}
\beta_1&=&\theta_1-\frac{(1-\epsilon)\theta_1}
{\sqrt{(1-\epsilon)\theta_1^2+(1+\epsilon)\theta_2^2}} , \\
\beta_2&=&\theta_2-\frac{(1+\epsilon)\theta_2}
{\sqrt{(1-\epsilon)\theta_1^2+(1+\epsilon)\theta_2^2}} ,  
\end{eqnarray}
where $\mbox{\boldmath$\beta$}=(\beta_1, \beta_2)$ and 
$\mbox{\boldmath$\theta$}=(\theta_1, \theta_2)$ denote 
the positions of the source and images, respectively. 

For simplicity, we introduce variables as 
$x\equiv\sqrt{1-\epsilon}\theta_1$, 
$y\equiv\sqrt{1+\epsilon}\theta_2$, 
$a\equiv\sqrt{1-\epsilon}\beta_1$ and 
$b\equiv\sqrt{1+\epsilon}\beta_2$, so that the lens equation 
can be rewritten as 
\begin{eqnarray}
a&=&x\left(1-\frac{1-\epsilon}{\sqrt{x^2+y^2}}\right) , 
\label{lenseq-sie1}
\\
b&=&y\left(1-\frac{1+\epsilon}{\sqrt{x^2+y^2}}\right) . 
\label{lenseq-sie2}
\end{eqnarray}

\subsection{Sources on the symmetry axes}
There are two symmetry axes in the ellipsoid, 
$a=0$ and $b=0$. 
We consider a source on the axis $a=0$. 
In this case, we can find analytic solutions for 
the lens equation as follows. 
We can make a replacement as $a \leftrightarrow b$, 
$x \leftrightarrow y$ and 
$1+\epsilon \leftrightarrow 1-\epsilon$ 
to obtain solutions for $b=0$.  

For $a=0$, Eq. ($\ref{lenseq-sie1}$) becomes 
\begin{equation}
x\left(1-\frac{1-\epsilon}{\sqrt{x^2+y^2}}\right)=0 , 
\end{equation}
which means 
\begin{equation}
x=0 , 
\label{sol-sie1} 
\end{equation}
or 
\begin{equation}
\frac{1-\epsilon}{\sqrt{x^2+y^2}}=1 . 
\label{sol-sie2}
\end{equation}

In the case of $x=0$, Eq. ($\ref{lenseq-sie2}$) becomes 
\begin{equation}
b-y=-(1+\epsilon)\frac{y}{|y|} , 
\end{equation}
which is solved as 
\begin{equation}
y=
\cases{
b-(1+\epsilon)
& for $b< -(1+\epsilon)$ , \cr
b\pm(1+\epsilon)
& for $-(1+\epsilon)\leq b\leq 1+\epsilon$ , \cr
b+(1+\epsilon)
& for $1+\epsilon < b$ . }
\label{sol-siey}
\end{equation} 
Here, we should pay attention to the cases of $b=\pm(1+\epsilon)$, 
since they produce a solution at the singularity 
$(x, y)=(0, 0)$ of the potential. 

Next, we consider the case of Eq. ($\ref{sol-sie2}$). 
Let us investigate the two cases, $\epsilon=0$ and 
$\epsilon\neq 0$ separately. 
For $\epsilon=0$, Eq. ($\ref{lenseq-sie2}$) means 
$b=0$, so that images become a ring as 
\begin{equation}
x^2+y^2=1 . 
\end{equation}
Below, we assume $\epsilon\neq 0$. 
Eliminating $\sqrt{x^2+y^2}$ from Eqs. ($\ref{lenseq-sie2}$) 
and ($\ref{sol-sie2}$), we obtain 
\begin{equation}
y=-\frac{(1-\epsilon)b}{2\epsilon} . 
\label{sol-siey2}
\end{equation}
Substituting this into $y$ in Eq. ($\ref{sol-sie2}$), 
we obtain 
\begin{equation}
x^2=\left(\frac{1-\epsilon}{2\epsilon}\right)^2(4\epsilon^2-b^2) , 
\end{equation}
which has real solutions if and only if $|b|\leq 2\epsilon$, 
\begin{equation}
x=\pm\frac{1-\epsilon}{2\epsilon}\sqrt{4\epsilon^2-b^2} . 
\label{sol-siex}
\end{equation}
Consequently, Eqs. ($\ref{sol-siey}$) and ($\ref{sol-siex}$) 
show that four, two or one images occur for $|b|<2\epsilon$, 
$2\epsilon<|b|<1+\epsilon$ or $1+\epsilon<|b|$, respectively. 

In the similar manner, we obtain the solutions for $b=0$. 
A point is that $2\epsilon<|a|<1-\epsilon$ can hold 
only for $\epsilon<1/3$.

\subsection{Off-axis sources}
Here, we consider off-axis sources ($a\neq 0$ and $b\neq 0$).
In this case, Eqs. ($\ref{lenseq-sie1}$) and 
($\ref{lenseq-sie2}$) show $x\neq 0$ and $y\neq 0$. 
Eliminating $\sqrt{x^2+y^2}$ from Eqs. ($\ref{lenseq-sie1}$) and 
($\ref{lenseq-sie2}$), we obtain 
\begin{equation}
y=\frac{(1-\epsilon)b x}{(1+\epsilon)a-2\epsilon x} , 
\label{siey}
\end{equation}
which determines $y$ uniquely for any given $x$. 
For finite $b$, Eq. ($\ref{lenseq-sie2}$) means that 
$y$ is also finite, so that 
$x\neq(1+\epsilon)a/2\epsilon$ from Eq. ($\ref{siey}$). 
Substituting Eq. ($\ref{siey}$) into Eq. ($\ref{lenseq-sie1}$), 
we obtain the fourth-order polynomial for $x$ as 
\begin{eqnarray} 
D(x)&\equiv&[(a-x)^2-(1-\epsilon)^2] 
[(1+\epsilon)a-2\epsilon x]^2 
\nonumber\\ 
&&+(1-\epsilon)^2b^2(a-x)^2 
\nonumber\\ 
&=& 0 , 
\label{fourth-eq}
\end{eqnarray} 
where we used $x\neq 0$.  

The number of real roots for a fourth-order equation is 
discussed by the discriminant $D_4$ (e.g. van der Waerden 1966), 
which becomes for Eq. ($\ref{fourth-eq}$) 
\begin{eqnarray}
D_4&=&-64a^2b^2\epsilon^2(1-\epsilon)^{12} 
\nonumber\\
&&\times[(a^2+b^2-4\epsilon^2)^3
+108a^2b^2\epsilon^2] . 
\end{eqnarray}
Namely, if 
\begin{equation}
\left(\frac{a^2}{4\epsilon^2}
+\frac{b^2}{4\epsilon^2}-1\right)^3 
+27\left(\frac{a^2}{4\epsilon^2}\right) 
\left(\frac{b^2}{4\epsilon^2}\right) < 0 , 
\label{D42}
\end{equation}
the number of real roots is either four or zero, which is 
determined as four by explicit solutions for on-axis sources. 
Otherwise, it is two. 
However, the number does not necessarily indicate 
that of images as shown below. 

Since $x\neq 0$ for off-axis sources, Eq. ($\ref{lenseq-sie1}$) 
is rewritten as 
\begin{equation}
\frac{a-x}{x}=-\frac{1-\epsilon}{\sqrt{x^2+y^2}} , 
\end{equation}
whose right-hand side is necessarily negative 
since $1-\epsilon>0$. 
As a result, we find that any solution of the lens equation 
must satisfy 
\begin{equation}
\frac{a-x}{x} < 0 . 
\label{condition-x}
\end{equation}
This implies that $x<0$ or $a<x$ for positive $a$, 
while $x<a$ or $0<x$ for negative $a$. 
It should be noted that Eq. ($\ref{condition-x}$) 
always holds in the limit of $a\to 0$. 

Let us investigate the number of roots in $(a-x)/x > 0$, 
namely an interval between $0$ and $a$. 
We find out 
\begin{eqnarray}
&&D(a)=-a^2(1-\epsilon)^4<0 , 
\\
&&D(0)=a^2(1-\epsilon^2)^2 
\left(
\frac{a^2}{(1-\epsilon)^2}+\frac{b^2}{(1+\epsilon)^2}-1
\right) . 
\end{eqnarray}
If $a$ and $b$ satisfy 
\begin{equation}
\frac{a^2}{(1-\epsilon)^2}+\frac{b^2}{(1+\epsilon)^2} \geq 1 , 
\label{D0}
\end{equation}
$D(0)$ is not negative, so that $D(x)=0$ has at least 
one root between $0$ and $a$. 
For $\epsilon<1/3$, Eq. ($\ref{D0}$) implies 
$a^2/4\epsilon^2+b^2/4\epsilon^2>1$, so that 
the left-hand side of Eq. ($\ref{D42}$) becomes positive.  
Hence, $D_4$ is negative, so that $D(x)=0$ has two roots. 
As a result, it has only one root in the interval, which means 
only one image appears. 
Unless Eq. ($\ref{D0}$) holds, $D(0)$ is negative, so that 
the number of roots for $D(x)=0$ for $(a-x)/x < 0$, 
namely that of images are four or two, respectively 
for $D_4>0$ or $<0$. 

The inner and outer caustics (Fig. 1) are given by 
\begin{equation}
\left(\frac{a^2}{4\epsilon^2}
+\frac{b^2}{4\epsilon^2}-1\right)^3 
+27\left(\frac{a^2}{4\epsilon^2}\right) 
\left(\frac{b^2}{4\epsilon^2}\right) = 0 , 
\label{caustic1}
\end{equation}
\begin{equation}
\frac{a^2}{(1-\epsilon)^2}+\frac{b^2}{(1+\epsilon)^2} = 1 . 
\label{caustic2}
\end{equation}
The inner caustic given by Eq. ($\ref{caustic1}$) is an asteroid 
which is parametrized as 
\begin{eqnarray}
&&a=2\epsilon\cos^3 t , 
\label{caustic1a}
\\
&&b=2\epsilon\sin^3 t , 
\label{caustic1b}
\end{eqnarray}
where $t\in [0, 2\pi)$. 

\begin{figure}
\centering
\includegraphics[width=8cm]{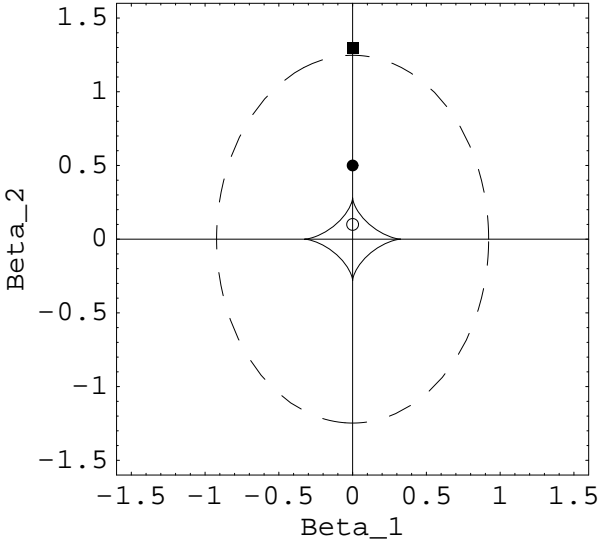}
      \caption{Caustics for a singular isothermal ellipsoidal lens 
in the physical coordinates $(\beta_1, \beta_2)$. 
For $\epsilon=0.15$,  the solid curve denotes the inner caustic, 
and the dashed one for the outer caustic. Sources locate at 
$(0, 0.1)$, $(0, 0.5)$ and $(0, 1.3)$, denoted by 
the circle, filled disk and square, respectively. 
}
\label{Fig1}
\end{figure}

\begin{figure}
\centering
\includegraphics[width=8cm]{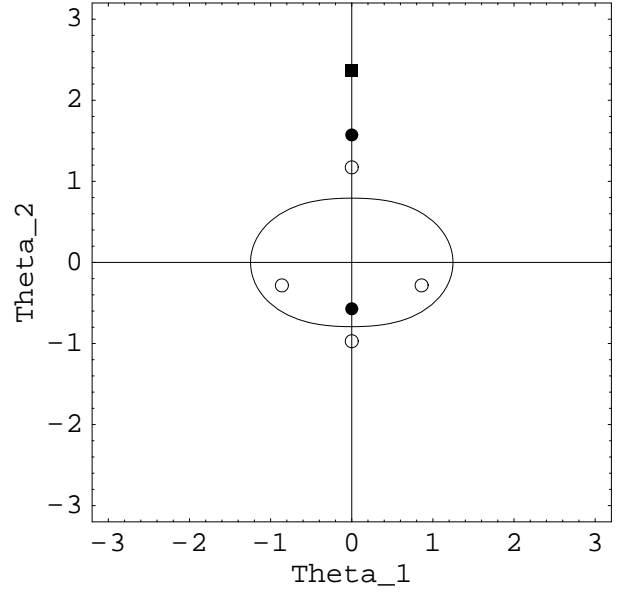}
      \caption{Critical curves for a singular isothermal ellipsoidal 
lens in the physical coordinates $(\theta_1, \theta_2)$. 
For $\epsilon=0.15$,  the solid curve denotes the outer critical curve 
which corresponds to the inner caustic. The origin corresponds to 
the outer caustic. The images correspond to the sources in Fig. 1. 
}
\label{Fig2}
\end{figure}

The critical curves on the lens plane correspond to the caustics 
on the source plane (Schneider et al. 1992). 
We introduce the polar coordinates as 
$(x, y)\equiv (\rho\cos\xi, \rho\sin\xi)$. 
By substituting Eqs. ($\ref{caustic1a}$) and ($\ref{caustic1b}$) 
with $t=-\xi$ into the lens equations ($\ref{lenseq-sie1}$) and 
($\ref{lenseq-sie2}$), we obtain the parametric representation of 
the critical curve as 
\begin{equation}
\rho=1+\epsilon\cos2\xi .
\label{critical1}
\end{equation}
In the similar manner to the outer caustic given by 
Eq. ($\ref{caustic2}$), we find 
\begin{equation}
\rho=0 , 
\label{critical2}
\end{equation} 
which is the origin in $(x, y)$ (Fig. 2).

\section{Lens Equation for a Cored Isothermal Ellipsoid} 
Let us consider a cored isothermal ellipsoidal lens 
with the angular core radius $c$. 
The lens equation is expressed as 
\begin{eqnarray}
a&=&x\left(1-\frac{1-\epsilon}{\sqrt{x^2+y^2+c^2}}\right) , 
\label{lenseq-cie1}
\\
b&=&y\left(1-\frac{1+\epsilon}{\sqrt{x^2+y^2+c^2}}\right) , 
\label{lenseq-cie2}
\end{eqnarray}
which are apparently similar to the set of 
Eqs. ($\ref{lenseq-sie1}$) and ($\ref{lenseq-sie2}$). 
However, there are differences in their algebraic properties 
as shown below. 

\subsection{Sources on the symmetry axes}
We consider a source on the axis $a=0$. 
In this case, we can find analytic solutions for 
the lens equation as follows. 
To consider the case of $b=0$, it is enough to make a 
replacement as $a \leftrightarrow b$, $x \leftrightarrow y$ 
and $1+\epsilon \leftrightarrow 1-\epsilon$. 

For $a=0$, Eq. ($\ref{lenseq-cie1}$) becomes 
\begin{equation}
x\left(1-\frac{1-\epsilon}{\sqrt{x^2+y^2+c^2}}\right)=0 , 
\end{equation}
which means 
\begin{equation}
x=0 , 
\label{sol-cie1} 
\end{equation}
or 
\begin{equation}
\frac{1-\epsilon}{\sqrt{x^2+y^2+c^2}}=1 . 
\label{sol-cie2}
\end{equation}

In the case of $x=0$, Eq. ($\ref{lenseq-cie2}$) 
becomes the fourth-order polynomial for $y$ as 
\begin{eqnarray}
E(y)&\equiv&y^4-2by^3+[b^2+c^2-(1+\epsilon)^2]y^2 
-2bc^2y+b^2c^2 
\nonumber\\ 
&&=0 . 
\label{fourth-eq2} 
\end{eqnarray}
Explicit solutions for a fourth-order equation take 
a lengthy form (e.g. van der Waerden 1966). 
Now, Eq. ($\ref{lenseq-cie2}$) implies 
\begin{equation}
\frac{b-y}{y} < 0 . 
\end{equation}
Using 
\begin{eqnarray}
&&E(0)=b^2c^2>0 , \\
&&E(b)=-b^2(1+\epsilon)^2<0 , 
\end{eqnarray}
we find that $E(y)$ has at least one zero point 
between $0$ and $b$. In other words, $E(y)=0$ has 
at most three roots for $(b-y)/y<0$. 

Next, we consider the case of Eq. ($\ref{sol-cie2}$). 
For $\epsilon=0$, Eq. ($\ref{lenseq-cie2}$) means $b=0$, 
so that a ring image appears at  
\begin{equation}
x^2+y^2+c^2=1 . 
\end{equation} 
We assume $\epsilon\neq 0$ in the following. 
Eliminating $\sqrt{x^2+y^2+c^2}$ from Eqs. ($\ref{lenseq-cie2}$) 
and ($\ref{sol-cie2}$), we obtain 
\begin{equation}
y=-\frac{(1-\epsilon)b}{2\epsilon} . 
\end{equation}
Substituting this into $y$ in Eq. ($\ref{sol-cie2}$), 
we obtain 
\begin{equation}
x^2=\left(\frac{1-\epsilon}{2\epsilon}\right)^2 
(4\epsilon^2-b^2)-c^2 , 
\end{equation}
which has the real solutions  
\begin{equation}
x=\pm\frac{1-\epsilon}{2\epsilon} \sqrt{4\epsilon^2 
\Bigl[1-\left(\frac{c}{1-\epsilon}\right)^2\Bigr]-b^2} ,  
\end{equation}
if and only if 
\begin{equation}
b^2\leq 4\epsilon^2 
\Bigl[1-\left(\frac{c}{1-\epsilon}\right)^2\Bigr] . 
\end{equation}

\subsection{Off-axis sources}
Here, we consider off-axis sources ($a\neq 0$ and $b\neq 0$).
In this case, Eqs. ($\ref{lenseq-cie1}$) and 
($\ref{lenseq-cie2}$) imply $x\neq 0$ and $y\neq 0$. 
Eliminating $\sqrt{x^2+y^2+c^2}$ from Eqs. ($\ref{lenseq-cie1}$) 
and ($\ref{lenseq-cie2}$), we obtain 
\begin{equation}
y=\frac{(1-\epsilon)b x}{(1+\epsilon)a-2\epsilon x} . 
\label{ciey} 
\end{equation}
Equation ($\ref{lenseq-cie2}$) shows that $y$ is finite 
for finite $b$, so that $x\neq(1+\epsilon)a/2\epsilon$ from 
Eq. ($\ref{ciey}$). 
Substituting Eq. ($\ref{ciey}$) into Eq. ($\ref{lenseq-cie1}$), 
we obtain the sixth-order polynomial for $x$ as 
\begin{eqnarray}
F(x) &\equiv&(a-x)^2
\nonumber\\
&&
\times\Bigl[(x^2+c^2)[(1+\epsilon)a-2\epsilon x]^2
+(1-\epsilon)^2b^2x^2\Bigr] 
\nonumber\\
&&-(1-\epsilon)^2x^2[(1+\epsilon)a-2\epsilon x]^2
\nonumber\\
&=& 0 . 
\label{sixth-eqx}
\end{eqnarray} 
This equation has at most six real solutions whose 
analytic expressions can not be given by algebraic manners 
(e.g. van der Waerden 1966). 
As shown below, however, six real roots {\it never} mean 
six images. 

In the same manner as for the singular isothermal ellipsoid, 
we obtain a condition for $x$ as 
\begin{equation}
\frac{a-x}{x} < 0 . 
\label{condition-x2}
\end{equation}

Let us prove that there exists a root between $0$ and $a$. 
We can find out 
\begin{eqnarray}
&&F(0)=a^4c^2(1+\epsilon)^2>0 , 
\\
&&F(a)=-a^4(1-\epsilon)^4<0 . 
\end{eqnarray}
Since $F(x)$ is a continuous function, $F(x)=0$ has 
at least one root in the interval. 
For $(a-x)/x < 0$, consequently, $F(x)=0$ has at most 
five solutions. Since the polynomial is sixth-order, 
the discriminant is not sufficient to determine the exact 
number of roots. Hence, the determination is beyond 
the scope of our paper.

\subsection{Polar coordinates}
Up to this point, we have used {2-dimensional} Cartesian 
coordinates: We must solve Eq. ($\ref{sixth-eqx}$) and 
choose appropriate roots which satisfy the inequality 
by Eq. ($\ref{condition-x2}$). 
Here, we adopt {\it 3-dimensional} polar coordinates 
to simplify the inequality, as shown below. 
By taking $c$ as a fictitious third dimension, we define 
\begin{eqnarray}
&&r=\sqrt{x^2+y^2+c^2}\geq c , 
\label{condition-r}
\\
&&\cos\Psi=\frac{c}{r} , 
\\
&&x=r\sin\Psi\cos\phi , 
\\
&&y=r\sin\Psi\sin\phi , 
\end{eqnarray}
where we can assume $\sin\Psi\geq 0$.
Then, Eqs. ($\ref{lenseq-cie1}$) and ($\ref{lenseq-cie2}$) 
are rewritten as 
\begin{eqnarray}
&&a=r\left(1-\frac{1-\epsilon}{r}\right)\sin\Psi\cos\phi , 
\label{lenseq-cie3}
\\
&&b=r\left(1-\frac{1+\epsilon}{r}\right)\sin\Psi\sin\phi . 
\label{lenseq-cie4}
\end{eqnarray}
We concentrate on off-axis sources ($a\neq 0$ and $b\neq 0$). 

Eliminating $\sin\Psi$ from Eqs. ($\ref{lenseq-cie3}$) and 
($\ref{lenseq-cie4}$), we obtain 
\begin{equation}
\tan\phi=\frac{b[r-(1-\epsilon)]}{a[r-(1+\epsilon)]} , 
\label{tan}
\end{equation}
which determines $\tan\phi$ uniquely for any given $r$. 
Equations ($\ref{lenseq-cie3}$) and ($\ref{lenseq-cie4}$) 
show that $r\neq 1\pm\epsilon$ for nonvanishing $a$ and $b$, 
since $x\neq 0$ and $y\neq 0$ mean $\cos\phi\neq 0$ and 
$\sin\phi\neq 0$. 
Hence, Eqs. ($\ref{lenseq-cie3}$) and ($\ref{lenseq-cie4}$) 
can be rewritten as 
\begin{eqnarray}
x&=&\frac{ar}{r-(1-\epsilon)} , 
\label{lenseq-cie5}
\\
y&=&\frac{br}{r-(1+\epsilon)} . 
\label{lenseq-cie6}
\end{eqnarray}
Substituting these into $r^2=x^2+y^2+c^2$, we obtain 
the sixth-order equation for $r$ as 
\begin{eqnarray}
G(r)&\equiv&(r^2-c^2)[r-(1-\epsilon)]^2[r-(1+\epsilon)]^2 
\nonumber\\
&&-a^2r^2[r-(1+\epsilon)]^2
-b^2r^2[r-(1+\epsilon)]^2 
\nonumber\\
&=& 0 , 
\label{sixth-eqr}
\end{eqnarray}
which has at most six real solutions. 
Let us show that there are at most five roots 
compatible with $r\geq c$. 
Using 
$G(c)=-a^2c^2[c-(1+\epsilon)]^2-b^2c^2[c-(1-\epsilon)]^2 < 0$ 
and $G(-\infty)=+\infty > 0$ for a continuous function $G(r)$, 
we find that $G(r)=0$ has at least one root for $r<c$. 
Consequently, it has at most five roots for $r\geq c$. 

We should note that $r$ and $\tan\phi$ are not enough to 
determine the location of images. A strategy for determining 
the location is as follows: 
First, we solve Eq. ($\ref{sixth-eqr}$) for $r\geq c$. 
Next, we substitute $r$ into Eqs. ($\ref{lenseq-cie5}$) 
and ($\ref{lenseq-cie6}$) to obtain the image position as 
$(x, y)$.

\section{Conclusion}
We have carefully reexamined the lens equation for a cored 
isothermal ellipsoid both in 2-dimensional Cartesian and 
3-dimensional polar coordinates. 
We have shown that the nonlinearly coupled equations are 
reduced to a single real sixth-order polynomial 
Eq. ($\ref{sixth-eqx}$) or ($\ref{sixth-eqr}$), which coincides 
with the fourth-order equation for a singular isothermal ellipsoid 
as the core radius approaches zero. 
For the singular case, explicit expressions of image positions 
for sources on the symmetry axis are given by Eqs. ($\ref{sol-sie1}$), 
($\ref{sol-siey}$), ($\ref{sol-siey2}$) and ($\ref{sol-siex}$). 
Furthermore, we have presented analytic criteria for determining 
the number of images, which correspond to the caustics 
given by Eqs. ($\ref{caustic1}$) and ($\ref{caustic2}$). 
Consequently, analytic expressions for the critical curves are 
given by Eqs. ($\ref{critical1}$) and ($\ref{critical2}$). 
We have shown for the cored case that a condition 
Eq. ($\ref{condition-x2}$) or ($\ref{condition-r}$) 
gives us physical solutions of the sixth-order polynomial, 
which are at most five images. 

The present formulation based on the one-dimensional equation 
($\ref{sixth-eqx}$) or ($\ref{sixth-eqr}$) enables us to study 
a cored isothermal ellipsoidal lens with considerable efficiency 
and accuracy, in comparison with previous two-dimensional 
treatments for which there are no well-established numerical methods 
(Press et al. 1988). 
Particularly for a source close to the caustics, the image 
position is unstable so that careful computations are needed. 
The amount of computations can be reduced by our approach. 
As a result, it must be powerful in rapid and accurate parameter 
fittings to observational data.

\begin{acknowledgements}
We would like to thank M. Bartelmann for carefully reading 
the manuscript and for a number of encouraging comments. 
H. A. would like to thank M. Bartelmann, L. Rezzolla and J. Miller 
for hospitality at the Max-Planck-Institut f\"ur Astrophysik 
and the Scuola Internazionale Superiore di Studi Avanzati, 
respectively, where a part of this work was done.
This work was supported in part by a Japanese Grant-in-Aid 
for Scientific Research from the Ministry of Education, 
No. 13740137 (H. A.) and the Sumitomo Foundation (H. A.).
\end{acknowledgements}

\end{document}